\documentclass[%
aps,%
prl,%
twocolumn,%preprint,%
showpacs,%
preprintnumbers,%
byrevtex,%
floatfix%
]{revtex4}

\usepackage{epsf}
\usepackage{ifthen}
\usepackage[usenames]{color}
\usepackage{amssymb}
\usepackage{amsfonts}
\usepackage{amsmath}
\usepackage[dvips]{graphicx}
\def\ii{\textrm{i}\,}

\def\ie{\textit{i.e.}}

\def\eg{\textit{e.g.}}

\begin{document}
%\date{December 11, 2008}

%\preprint{\texttt{cond-mat/04?????}}
\title{%
Charge transport through bio-molecular wires in a solvent: Bridging molecular dynamics and model Hamiltonian approaches
}

\author{R. Guti{\'e}rrez$^{1}$}
\author{R. Caetano$^{1,2}$}
\author{B. Woiczikowski$^{3}$}
\author{T. Kubar$^{3}$}
\author{M. Elstner$^{3}$}
\author{G. Cuniberti$^{1}$}

\affiliation{%
$^{1}$ Institute for Materials Science and Max Bergmann Center of Biomaterials, Dresden University of Technology, D-01062, Dresden, Germany \\
$^{2}$ Instituto de Fisica, Universidade Federal de Alagoas, Maceio, AL 57072-970, Brazil\\
$^{3}$ Institute for Physical and Theoretical Chemistry, Technical University Braunschweig, D-38106, Braunschweig, Germany}

\title{Charge transport through bio-molecular wires in a solvent: Bridging molecular dynamics and model Hamiltonian approaches}

\begin{abstract}
We present a hybrid method based on a combination of quantum/classical molecular dynamics (MD) simulations and a model Hamiltonian approach to describe charge transport through bio-molecular wires with variable lengths in presence of a solvent. The core of our approach consists in a mapping of the bio-molecular electronic structure, as obtained from density-functional based tight-binding calculations of molecular structures along MD trajectories, onto a low dimensional model Hamiltonian including the coupling to a dissipative bosonic environment. The latter encodes fluctuation effects arising from the solvent and from the molecular conformational dynamics. We apply this approach to the case of pG-pC and pA-pT DNA oligomers as paradigmatic cases and show that the DNA conformational fluctuations are essential in determining and supporting charge transport.
\end{abstract}
\maketitle

Can a DNA molecular wire support an electrical current? The variety of partially contradictory experimental results obtained in the past years~\cite{porath00,schuster04,ron2006,storm01,yoo01,tao04,cohen05} has hinted not only at the difficulties encountered to carry out well-controlled transport measurements, but also at the strong sensitivity of charge migration to intrinsic (base-pair sequence, internal vibrations) and extrinsic (solvent fluctuations, molecule-electrode contact) factors. Recently~\cite{tao04,cohen05}, two experimental groups have measured similar  high electrical currents on the order of 50-100 nA despite the fact that the electrically probed base sequences and lengths  were rather different.
In spite of considerable theoretical research, the dominant mechanisms for charge transport through DNA wires have not been, however, fully elucidated, see \eg, Ref.~\cite{tapash07} for a recent review.  Electron transfer experiments~\cite{wan99,O'NeillM.A._ja0455897} as well as related theoretical studies~\cite{jortner02,bruinsma00,GrozemaF_ja001497f,mallajosyula:176805,troisi02,CramerT._jp071618z,barnett01, gmc05a,gmc05b,gutierrez06,schmidt:115125,schmidt:165337,GrozemaFerdinandC._ja078162j,voityuk04} have clearly pointed out the crucial role of  dynamical fluctuations in favouring or hindering hole transfer. We may thus expect that this may also be the case for {\textit{charge transport}}. Studies based on model Hamiltonian formulations describing disordered sequences~\cite{roche03,klotsa05} or the coupling to dynamical degrees of freedom~\cite{gmc05a,gmc05b,gutierrez06,schmidt:115125,schmidt:165337} involve many free parameters which are in general difficult to determine for realistic situations.
First-principle calculations~\cite{difelice02,bixon00,voityuk:064505,mehrez05,dima2008,beratan2008} performed on static structures  provide, on the other hand, orders of magnitude for the electronic coupling parameters but can hardly deal  with  the coupling to dynamical degrees of freedom. The inclusion of dynamical effects in quantum transport calculations has only been addressed in few especial  cases~\cite{starikov07,mallajosyula:176805,hennig04,CramerT._jp071618z} in a systematic way.  Thus, a general approach able to combine dynamical information drawn from a realistic description of bio-molecular conformational dynamics with a treatment of quantum transport is highly desirable.

%%%%%%%%%%%%%%%%%%%%%%%%%%%%%%% FIGURE 1 %%%%%%%%%%%%%%%%%%%%%%%%%%%%%%%%%%%%%%%%%%%%%%%%%%%%%
\begin{figure}[b]
\centerline{
\epsfclipon
\includegraphics[width=.99\linewidth]{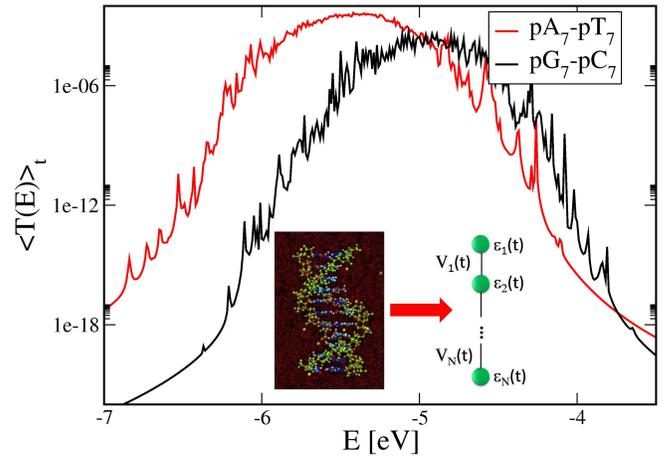}%
\epsfclipoff
}
\caption{\label{fig:adiabatic}%
Typical time-averaged transmission function $\left\langle T(E)\right\rangle_{t}$ for pG-pC and pA-pT wires containing seven base pairs each. The time series length was 100 ps and the average was performed every 5 ps. Inset: A snapshot of a pG-pC oligomer with 11 base pairs in a solvent drawn from the MD simulation. To avoid spurious boundary effects, only the innermost seven base pairs are used. Via the fragment orbital method, the electronic structure at each snapshot is mapped onto an effective low dimensional model Hamiltonian with time-dependent electronic parameters.
}
\end{figure}
In this Letter, we present a study of charge transport through bio-molecular wires  with different lengths by using a hybrid approach based on a mapping of the  time-fluctuating  electronic structure along a molecular dynamics (MD) trajectory onto a low-dimensional model Hamiltonian. Charge transport will be studied for an effective model describing  the coupling of the  electronic system  to a bosonic bath which comprises internal vibrations and solvent effects. The bath thus encodes dynamical information drawn from the MD simulations. Our treatment allows (i) the determination of electronic coupling parameters under {\textit {realistic}} conditions, and (ii) the calculation of the bath spectral density from time series generated during the MD simulation.
%~\cite{PhysRevE.65.031919} 
In this way, it does not contain any free parameters describing the molecular electronic structure or the coupling to the structural fluctuations.  We show, by considering as paradigmatic cases pG-pC and pA-pT oligomers, that the electrical transport properties in such bio-molecular systems are strongly dominated by the conformational dynamics. Despite some limitations of the model Hamiltonian approach related to the used approximations (see later), we nevertheless stress that the range of applicability of our method is not limited to DNA molecular wires; indeed, it provides a solid basis for the parameter-free inclusion of dynamical effects in a model-based treatment of quantum transport as well as for a multi-scale approach to the description of the electronic properties of bio-molecules and their response to external fields. 
%%%%%%%%%%%%%%%%%%% FIGURE 2 %%%%%%%%%%%%%%%%%%%%%%%%%%%%%%%%%%%%%%%%%%%%%%%%%%%%%%%%%
\begin{figure}[t]
\centerline{
\epsfclipon
\includegraphics[width=.99\linewidth]{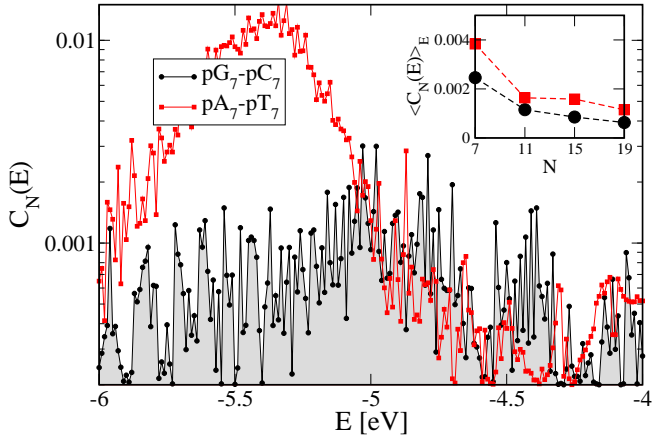}%
\epsfclipoff
}
\caption{\label{fig:fig2}%
Coherence parameter $C_{N}(E)$ which provides a qualitative measure for the importance of dynamical effects; $C_{N}(E)\ll 1$ hints at the dominant role of fluctuations in determining the transport properties. Notice, however, that this parameter can only give reliable information when $\left\langle T(E)\right\rangle_{t}\neq 0$, \ie, within the energy window belonging to the spectral support of $\left\langle T(E)\right\rangle_{t}$. The inset shows the energy averaged $\left\langle C_{N}(E)\right\rangle_{E}$  as a function of the number of base pairs in the DNA chain. 
}
\end{figure}
Our approach exploits a fragment orbital description~\cite{SGG2005} of the bio-molecules, which allows for an efficient and well controlled coarse graining of the electronic structure problem~\cite{markus1,markus2}. A combination of quantum mechanics/molecular mechanics  methods (to describe solvent effects), MD simulations, and a parametrized density-functional tight-binding methodology~\cite{dftb} has been used to extract the relevant electronic information in the form of time series. This leads to a time-dependent Hamiltonian:
%\begin{eqnarray}
 $H=\sum_{j} \epsilon_{j}(t) d^{\dagger}_{j}d_{j} + \sum_{j} V_{j,j+1}(t)\,(d^{\dagger}_{j}d_{j+1} + {\textrm {h.c.}})$,
%\end{eqnarray}
where both $\epsilon_{j}(t)$ and  $ V_{j,j+1}(t)$ are random variables as a function of the simulation time. These parameters describe, respectively, the effective ionization energy of a single base-pair $-$which defines a fragment in our calculations$-$ and the coupling between nearest-neighbor fragments. We have approached the transport problem from two complementary perspectives.
% %%%%%%%%%%%%%%%%%%%%%%%%%%%%%%% FIGURE 3 %%%%%%%%%%%%%%%%%%%%%%%%%%%%%%%%%%%%%%%%%%%%%%%%%%%%%
\begin{figure}[t]
\centerline{
\epsfclipon
\includegraphics[width=.99\linewidth]{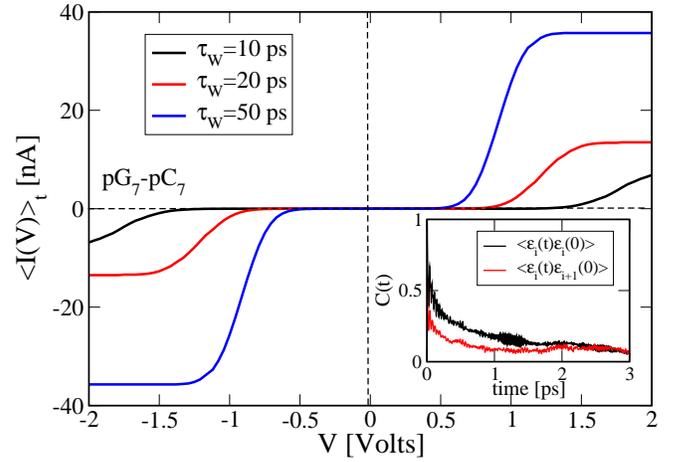}%
\epsfclipoff
}
\caption{\label{fig:current}%
The time-averaged current $\left\langle I(V)\right\rangle_{t}$ for a pG$_{7}$-pC$_{7}$ oligomer. Results are shown for three different choices of the time frame $\tau_{\textrm{W}}=$5, 20, and 50 ps (for a definition see the main text). Inset: Auto-correlation function of the onsite energy fluctuations $\left\langle \epsilon_{i}(t) \epsilon_{i}(0)\right\rangle$ and nearest-neighbor cross-correlation $\left\langle \epsilon_{i}(t) \epsilon_{i+1}(0)\right\rangle$. 
}
\end{figure}

 \paragraph{Time averaging and dynamical fluctuations $-$}  In Fig.~1 we show the time-averaged transmission function 
$\left\langle T(E)\right\rangle_{t}$ for  pG-pC and pA-pT wires containing seven base pairs~\cite{comment1}. These calculations have been carried out for a $T_{\textrm{MD}}=30$ ns long MD simulation with a time step of 1 fs. The first point to note is the apparently higher transmission of pA-pT compared to that of pG-pC. This is just the opposite of what a purely static calculation would yield. This fact represents a first hint at the importance of dynamical effects in determining charge propagation. The fragmented structure of the spectrum is simply mirroring the broad distribution of onsite energies induced by the dynamical disorder. We have further defined a coherence parameter (CP) for a given chain length $N$ as $C_{N}(E)=(1+\sigma_{T}(E)/\left\langle T(E)\right\rangle_{t}^{2})^{-1}$, which can provide a quantitative measure for the role of structural fluctuations: $C_{N}(E)\ll 1$ can indicate the dominance of the conformational dynamics. Hereby, $\sigma_{T}(E)=\left\langle [T(E)-\left\langle T(E) \right\rangle_{t}]^{2}\right\rangle_{t}$ is the mean-square deviation of the transmission  and the brackets always indicate time averaging. Obviously, the CP has only a clear meaning in the spectral region  where the transmission function itself is non-negligible (spectral support). The CP is shown in Fig~2 for seven base pairs of pA-pT and pG-pC, from where it is seen that (i) transport is dominated by the bio-molecular dynamics, $C_{N}(E)\ll 1$, and (ii) the CP for pG-pC is roughly one order of magnitude smaller (within the spectral support domain) than for pA-pT, reflecting the fact that the latter system seems to be less affected by dynamical disorder. The inset of the same figure displays the energy-averaged CP for four different numbers of base pairs. Longer chains are clearly more affected by dynamical disorder than shorter chains, independent of the base sequence.  In a second step, we have investigated the dependence of the time-average current $\left\langle I(V)\right\rangle_{t}$ on the averaging procedure, \ie,  calculating a set of partial currents $I_{l}(V,\tau_{\textrm{W}})$ obtained upon averaging of the electronic parameters over time windows of length  $\tau_{\textrm{W}}=n_{{d}}\delta t$ along the time series, where $\delta t= 1$ ps is the time step at which molecular conformations were extracted along the MD trajectory.  The index $\ell=1,\cdots,{\textrm {int}}[T_{\textrm{MD}}/\tau_{\textrm{W}}]=L$ labels the number of time frames once $n_{d}$ has been fixed. The total current is thus given by $\left\langle I(V)\right\rangle_{t} = (1/L)\sum_{\ell} I_{\ell}(V,\tau_{\textrm{W}})$. The different sizes of the time windows (different values of $\tau_{\textrm{W}}$) are mirroring in a {\textit {phenomenological}} way differences in electronic time scales (an information not provided by the MD simulations); a charge will explore different fluctuating environments in dependence of $\tau_{{\textrm{W}}}$ and thus the total current must be affected by this fact. Hence \eg, $\tau_{\textrm{W}}\ll \omega^{-1}$, with  $\omega^{-1}$ being some typical time scale for dynamic fluctuations, would correspond to the non-adiabatic limit where a time-averaged atomic frame is felt by the charge, while the opposite limit $\tau_{\textrm{W}}\gg \omega^{-1}$ defines the adiabatic regime, where instantaneous atomic configurations are ``seen''. Of course, this provides only a qualitative picture, since the DNA structural fluctuations  involve many different time scales making the effective interaction of a charge with different degrees of freedom very complex.  In Fig.~3, we show the time-averaged current for a fixed number of base pairs and three different values of the time frame: $\tau_{\textrm{W}}=$5, 20, and 50 ps. We see a slow increase of the current with increasing $\tau_{\textrm{W}}$, since  a moving charge will effectively sample an increasingly larger number of realizations of $\epsilon_{j}(t)$ and $ V_{j,j+1}(t)$. We remark that the current calculation using a Landauer-like expression is only meaningfull near the adiabatic limit; in the strongly non-adiabtic regime golden-rule like expressions should be used~\cite{weissbook}.
% %%%%%%%%%%%%%%%%%%%%%%%%%%%%%%%%%%%%%%  Figure4  %%%%%%%%%%%%%%%%%%%%%%%%%%%%%%%%%%%%%%%%
\begin{figure}[t]
\centerline{
\epsfclipon
\includegraphics[width=.99\linewidth]{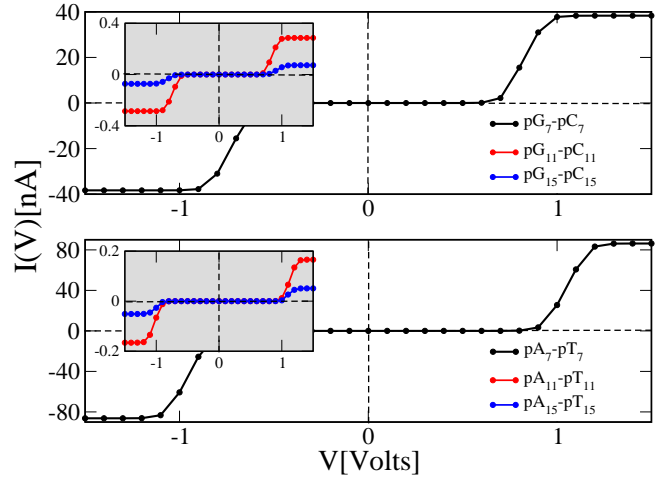}%
\epsfclipoff
}
\caption{\label{fig:curr}%
$I$-$V$ characteristics of pG-pC (upper panel) and pA-pT (lower panel) oligomers with three different numbers of base-pairs $N=7,11,15$ calculated with the effective charge-bath model Hamiltonian, see Eq.~(1). In all cases the electrode-molecule coupling parameter was chosen  $\Gamma=5$ $ meV$.~\cite{comment2} Though for the shortest oligomer pA-pT has a larger current than pG-pC, this behaviour is weakened with increasing length. Possible reasons for this behavior are mentioned in the text. 
% The solid lines in the upper panel correspond to the computed current while setting the charge-bath coupling to zero in Eq.~1. 
% Slight changes of the gap are related to the polaron shift $\left\langle\epsilon_{j}\right\rangle_{t}\rightarrow \left\langle \epsilon_{j}\right\rangle_{t}- \int_{0}^{\infty} d\omega J(\omega)/\omega$ of the onsite energies in Eq.~1.
}
\end{figure}

\paragraph{Effective charge-bath model $-$} As a complementary way to deal with charge transport $-$allowing for a flexible treatment of different transport mechanisms while still relying on a realistic description of the bio-molecular dynamics$-$ an effective model has been formulated describing the electronic system  coupled  to a fluctuating environment (bosonic bath):
\begin{eqnarray}
\label{eq1}
 H&=&\sum_{j} \left\langle \epsilon_{j}\right\rangle_{t}  d^{\dagger}_{j}d_{j} + \sum_{j} \left\langle {V}_{j,j+1}\right\rangle_{t}  \,(d^{\dagger}_{j}d_{j+1} + {\textrm {h.c.}}) \nonumber \\
&+&\sum_{s,j} \lambda_{sj} d^{\dagger}_{j}d_{j} (B_{s}+B^{\dagger}_{s})+ \sum_{s} \Omega_{s} B^{\dagger}_{s} B_{s}.
%&+&\sum_{{\bf k},\alpha} (t_{{\bf k},\alpha} c^{\dagger}_{{\bf k},\alpha} d_{1} + {\textrm h.c.})+ \sum_{{\bf k},\alpha}  \epsilon_{{\bf k},\alpha} c^{\dagger}_{{\bf k},\alpha} c_{{\bf k},\alpha}.
\end{eqnarray}
Here, the time averages of the electronic parameters $ \left\langle \epsilon_{j}\right\rangle_{t},\left\langle {V}_{j,j+1}\right\rangle_{t}$ have been splitted off, \eg,~$\epsilon_{j}(t)=\left\langle \epsilon_{j}\right\rangle_{t}+ \delta\epsilon_{j}(t)$. Some  approximations are involved by the formulation of this model: (i) only {\textit{local}} energy fluctuations   are considered and included in the  bath (last term of Eq.~(1)). (ii)  $\lambda_{sj}$ ($s$ numbers the bath modes) depends in general on the site $j$. This will be partially taken into account and is reflected in a renormalization of the average hopping $\left\langle {V}_{j,j+1}\right\rangle_{t}$; (iii) no fluctuations in the coupling parameters $V_{j,j+1}(t)$ are considered. 
The  bath will be characterized by a site-averaged  spectral density $J(\omega)=\left\langle\delta\epsilon^{2}(0)\right\rangle(2/\pi\hbar) \tanh{(\hbar\omega/k_{\textrm B} T)} \int_{0}^{\infty} dt\, \cos{(\omega t)} \, C(t)$,  with $C(t)=(1/N)\sum_{j}\left\langle\delta\epsilon_{j}(t)\delta\epsilon_{j}(0) \right\rangle$ being the (site-averaged) autocorrelation function of the onsite energy fluctuations.
% ~\cite{PhysRevE.65.031919}  
Using the model of Eq.~(\ref{eq1}), the electrical current through pG-pC and pA-pT oligomers containing $N$=7, 11, and 15 base pairs was computed~\cite{comment2}. The Fermi level was fixed in each case outside the spectral support of $T(E)$ to obtain a zero-current gap, which is thus arbitrary in these calculations. In Fig.~4, where  the $I$-$V$ characteristics of the different sequences and lengths are shown, we observe that apart from the shortest ($N=7$) oligomer, the  current for pG-pC is somewhat larger than for pA-pT. These features are possibly associated to two factors. Firstly, the neglect of non-local onsite energy fluctuations: fluctuations between neighboring sites decay faster on sub-$ps$ time scales, see the inset of Fig.~4, but can nevertheless induce  non-vanishing correlations over few base pairs~\cite{markus2} thus modifying the electrical response of the system. Secondly, the use of an averaged spectral density effectively makes the coupling of all electronic sites to the bath very similar. Since fluctuations become more important with increasing length, this approximation may become problematic. 
%Further,  the current without coupling to the bath is reduced for all lengths, see the upper panel of Fig.~\ref{fig:curr}; this  points out at the role of conformational fluctuations in favouring charge transport.
 In spite of these limitations, Eq.~(\ref{eq1}) provides a reasonable starting point to bridge MD simulations with charge transport models and, more important, offers the possibility of systematically improving the model Hamiltonian approach.

In  conclusion, we have presented a combined  molecular dynamics/model Hamiltonian approach which allows for a very flexible treatment of charge transport through bio-molecular systems taking into account dynamical disorder. The method can allow, as illustrated in the special case of DNA wires, the straightforward study of the base-sequence and length dependence of the electrical response of such systems. The results presented here strongly support the view that charge transport through DNA wires is dominated by  conformational fluctuations. In this sense, transport approaches based on  band-like coherent transport or on purely static structures can not yield a realistic description of charge motion in such highly dynamical systems. Finally, we would like to stress that our approach can  be applied as well to investigate the interplay of charge transport/transfer and conformational dynamics in other complex bio-molecular systems. This flexibility relies on the fact that the degree of coarse-graining $-$leading to the formulation of effective tight-binding models$-$ can be ``tuned'' by an appropriate re-definition of the fragment orbitals while still retaining the relevant dynamical information. 

The authors thank Stanislaw Avdoshenko and Jewgeni Starikov for useful discussions. This work has been  partially supported by the Deutsche Forschungsgemeinschaft (DFG) under contracts CU 44/5-2 and CU 44/3-2 and by the South Korea Ministry of Education, Science and Technology Program "World Class University" under contract R31-2008-000-10100-0.

\providecommand{\refin}[1]{\\ \textbf{Referenced in:} #1}


\begin{thebibliography}{10}
\bibitem{comment1}
 The current is calculated at this stage using the expression $\left\langle
  I(V)\right\rangle_{t}=(2e/h)\int dE (f_{\textrm{L}}(E)-f_{\textrm{R}}(E))
  \left\langle T(E)\right\rangle_{t}$, where
  $T(E)=\Gamma_{\textrm{L}}\Gamma_{\textrm{R}}|G_{1N}(E)|^{2}$ is the
  transmission probability for a linear chain of length $N$. The coupling to
  the $\alpha$-electrode $\Gamma_{\alpha}$ is treated as a free,
  energy-independent parameter (wide-band approximation). For the sake of
  simplicity we further assume
  $\Gamma_{\textrm{L}}=\Gamma_{\textrm{R}}=\Gamma$. Though our approach allows
  to determine the influence of the dynamics onto the molecule-electrode
  coupling, this increases the complexity of our analysis and will be left for
  future investigations.
\end{thebibliography}
\end{document}